\documentclass[journal,article,pdftex,moreauthors]{Definitions/mdpi} 
\def\trento{T$_{\rm R}$ENTo \hspace{0.07cm}}
\firstpage{1} 
\pubvolume{1}
\issuenum{1}
\articlenumber{0}
\pubyear{2023}
\copyrightyear{2023}
\datereceived{ } 
\daterevised{ } % Comment out if no revised date
\dateaccepted{ } 
\datepublished{ } 
\hreflink{https://doi.org/} % If needed use \linebreak
\Title{Momentum anisotropy generation in a hybrid approach$^\dagger$}

\TitleCitation{Momentum anisotropy generation in a hybrid approach}

\Author{Niklas G\"otz$^{1,2}$\orcidA{}*, Lucas Constantin$^{1}$  and Hannah Elfner$^{3,1,2,4}$}

\AuthorNames{Niklas G\"otz, Lucas Constantin and Hannah Elfner}

\AuthorCitation{G\"otz, N.; Constantin, L.; Elfner, H.}
\address{%
$^{1}$ \quad Goethe University Frankfurt, Department of Physics, Institute for Theoretical Physics, Frankfurt, Germany\\
$^{2}$ \quad Frankfurt Institute for Advanced Studies, Ruth-Moufang-Strasse 1, 60438 Frankfurt am Main, Germany \\
$^{3}$ \quad GSI Helmholtzzentrum f\"ur Schwerionenforschung, Planckstr. 1, 64291
Darmstadt, Germany \\
$^{4}$ \quad Helmholtz Research Academy Hesse for FAIR (HFHF), GSI Helmholtz Center, Campus Frankfurt, Max-von-Laue-Straße 12, 60438 Frankfurt am Main, Germany \\}

\corres{Correspondence: goetz@fias.uni-frankfurt.de}

\firstnote{Presented at Hot Quarks 2022-Workshop for Young Scientists on the Physics of Ultrarelativistic Nucleus-Nucleus Collisions, Estes Park, USA, 11–17 October 2022.} 
\abstract{Anisotropic flow emerges in all three of hybrid approaches: initial conditions, viscous relativistic hydrodynamics as well as hadronic transport. Previous works focus mainly on a constant or temperature dependent shear viscosity $\eta/s$. Here instead, we study qualitatively the effect of a generalized $\eta/s(T,\mu_B)$ in the hybrid approach SMASH-vHLLE-hybrid. The parameterization takes into account the constraints of matching to the transport coefficients in the hadronic phase, as well as of recent Bayesian analysis results. We compare the effect of the different parameterizations in the intermediate energy region of  $\sqrt{s_{NN}}$=7.7 -- 39.0 GeV. We observe that using the energy density dependent parameterization decreases the effect of the point of particlization. In addition, we quantify the uncertainty due to different initial state profiles, including the SMASH initial conditions as well as T\raisebox{-.5ex}{R}ENTo and IP-Glasma profiles. It can be shown that the initial state transverse momentum impacts final state momentum anisotropy.}


\begin{document}

%%%%%%%%%%%%%%%%%%%%%%%%%%%%%%%%%%%%%%%%%%

\section{Introduction}
There is an increasing interest in the theoretical description of heavy-ion collisions at finite baryon densities using hybrid approaches, which combine hadronic transport for the simulation of the early and late stages with relativistic hydrodynamics for the hot and dense stage. The inputs of the hydrodynamic evolution, especially the initial condition as well as the shear viscosity, determine the final state momentum anisotropy.

Many theoretical predictions support a non-constant shear viscosity over entropy ratio $\eta/s$, with a minimum close to the phase transition \cite{JETSCAPE:2020mzn,Ghiglieri:2018dib,Auvinen:2020mpc}. Additionally, there also exists theoretical predictions of a dependence on the net baryochemical potential \cite{Kadam_2015,Demir_2009}. Existing studies focus mainly on a temperature dependence or even only a constant effective shear viscosity. Therefore, the following investigates the effects of including both a temperature and a baryochemical potential dependence in the shear viscosity on the evolution and observables of a hybrid simulation. A more in-depth examination of the dependence of the shear viscosity on the net baryon chemical potential can be found in \cite{Gotz:2022naz}.
Additionally, there exists significant theoretical uncertainty about the initial conditions of the hydrodynamic evolution. Due to the extremely short lifetime, the initial state of heavy ion collisions is experimentally not accessible, leading to a variety of initial state models to exist \cite{Petersen:2008dd,Schenke:2012wb,Moreland:2014oya,Schafer:2021csj}.  This limits the predictive strength of a Bayesian inference based on a singular initial state model. We compare in the following the effect of changing initial condition models in our framework.

%%%%%%%%%%%%%%%%%%%%%%%%%%%%%%%%%%%%%%%%%%
\section{Setup}
The following results were calculated from the hybrid approach {SMASH-vHLLE-hybrid} \cite{hybridurl, Schafer:2021csj}. In order to create the SMASH initial condition, the hadronic transport approach is run until all particles have reached an eigentime equal to the geometrical overlap time of the two nuclei. The transition to the late stage rescattering is governed by the switching energy density $\epsilon_{sw}$, which is a free technical parameter and only controls when the hydrodynamic or the hadronic transport model is applied. The choice of the shear viscosity will be discussed in the following.

In contrast to adding terms proportional to $\mu_B$ to $\eta/s (T)$, parameterizing instead in the local rest frame energy density $\epsilon$ and the net baryon number $\rho$ has the advantage that these are evolved throughout hydrodynamic simulation, and therefore reduces the dependency of results on the choice of the equation of state.

We restrict ourselves to a linear dependence of the shear viscosity both in the region of high and low energy densities, with a minimum near transition. This already implies an implicit $\mu_B$-dependence. Additionally,  we study a linear term proportional to the net baryon number density $\rho$. This results in the following functional form:
\begin{equation}
    \eta/s (\epsilon, \rho) = \max\left(0,  (\eta/s)_{\text{kink}}  + \begin{cases}
S_{\epsilon, H}(\epsilon - \epsilon_{\text{kink}}) & \epsilon < \epsilon_{\text{kink}}\\
S_{\epsilon, Q}(\epsilon - \epsilon_{\text{kink}})  & \epsilon > \epsilon_{\text{kink}}
\end{cases} \right)
\end{equation}
According to experimental results \cite{Cleymans_2006}, we set the position of the minimum at vanishing net baryochemical potential to 1 GeV/fm$^3$, and the value the shear viscosity at this point is set to the KSS-bound \cite{Kovtun:2004de}. For high energy densities, the steepness is motivated by matching pQCD results \cite{Ghiglieri:2018dib}, whereas at low energy density we match to the shear viscosity extracted from box calculations in {SMASH} at the particlization temperature.

This parameterization is compared to other existing choices for constant or temperature-dependent $\eta/s$. The first choice is a constant value for $\eta/s$, with values depending on collision energies \cite{Schafer:2021csj}. 
The second choice is representative for results from Bayesian analysis \cite{JETSCAPE:2020mzn}.

%%%%%%%%%%%%%%%%%%%%%%%%%%%%%%%%%%%%%%%%%%
\section{Results}
In the following, we investigate the impact on the final state momentum anisotropy when changing both the shear viscosity in the hydrodynamic evolution as well as the initial condition of the hydrodynamic stage.
\subsection{Effect of baryochemical potential dependent shear viscosity on elliptic flow}
\begin{figure}
\centering
    \includegraphics[scale=1.15]{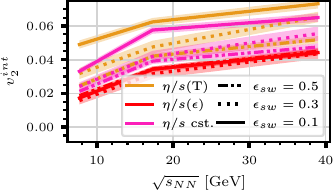}
    \caption{Integrated event plane elliptic flow of charged hadrons at midrapidity ($|y|<0.5$) for different parameterization strategies and values of $\epsilon_{\text{sw}}$.}
    \label{fig:main}
\end{figure}
Anisotropic flow is generated both in the hydrodrynamic as well as the transport stage. Whereas the shear viscosity, which has a strong impact on momentum anisotropies, is an input parameter in the hydrodynamic stage, it is not readily accessible in the non-equilibrium evolution. As the change of descriptions is governed by $\epsilon_{\text{sw}}$, information about the shear viscosity in the transport stage can be gained by varying this parameter.
The effect of a different choice of the switching energy density on the integrated elliptic flow of charged hadrons is plotted in Figure \ref{fig:main} at three investigated values for $\epsilon_{\text{sw}}$ for the default choice of $\eta/s$ as well as the parameterizations in $T$ and $\epsilon$. We see that the lines for constant $\eta/s$ and $\eta/s(T)$ show significant changes when varying $\epsilon_{\text{sw}}$, as the flow increases when reducing $\epsilon_{\text{sw}}$. In contrast, for $\eta/s(\epsilon)$, the lines stay close to each other, which in turn means that the flow is, for this range of $\epsilon_{\text{sw}}$, almost independent of $\eta/s(\epsilon)$.

By increasing $\epsilon_{\text{sw}}$, regions which were evolved in the hydrodynamic evolution are then evolved in hadronic transport, where the shear viscosity is not directly accessible. This means that the independence of the integrated flow from the value of $\epsilon_{\text{sw}}$ for $\eta/s(\epsilon)$ is a strong sign that $\eta/s(\epsilon)$ approximates the shear viscosity in the non-equilibrium hadronic transport stage. 

\subsection{Effect of exchanging initial conditions}
\begin{figure*}
    \centering
    \includegraphics[scale=0.85]{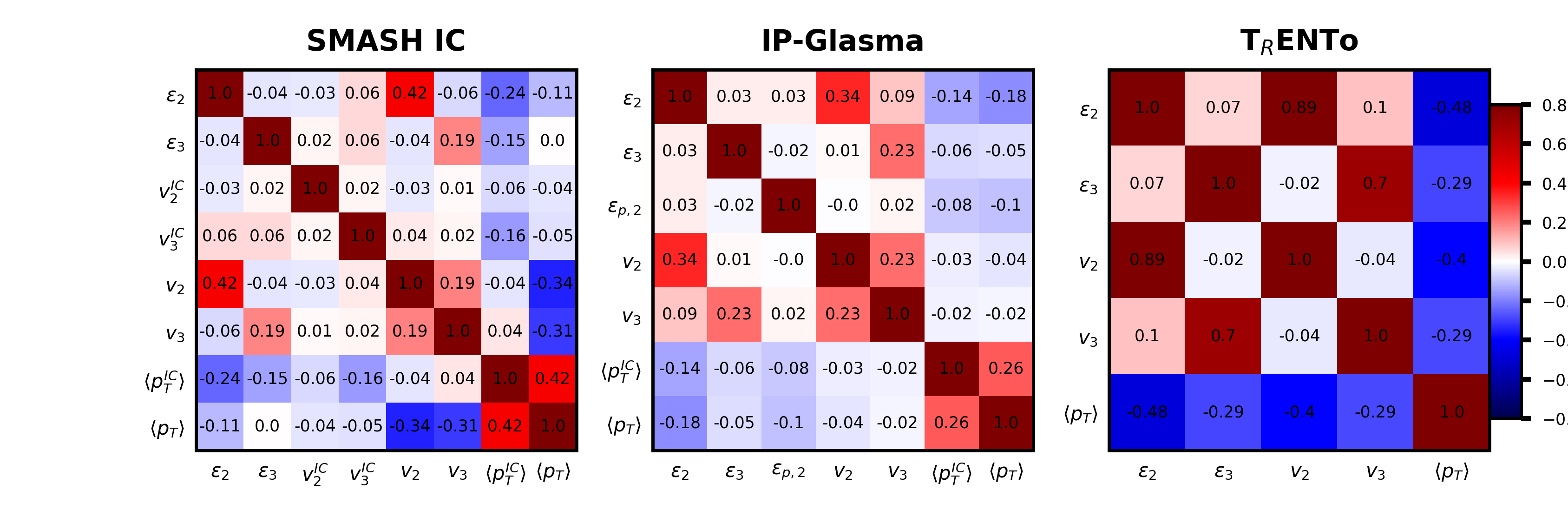}
    \caption{Pearson correlation matrix for all three initial state models, at 20-30\% centrality.}
    \label{fig:corr}
\end{figure*}

Anisotropic flow is commonly seen as a reaction to initial state eccentricities of nuclear matter, with the reaction mostly governed by the value of the shear viscosity. Using 750 events for Au-Au collisions at $\sqrt{s_{NN}}$= 200 GeV of SMASH initial conditions, IP-Glasma \cite{Schenke:2012wb} and \trento \cite{Moreland:2014oya} each, we test this hypothesis while employing the energy-density-dependent shear viscosity parameterization.
\begin{figure}
    \centering
    \includegraphics[scale=0.65]{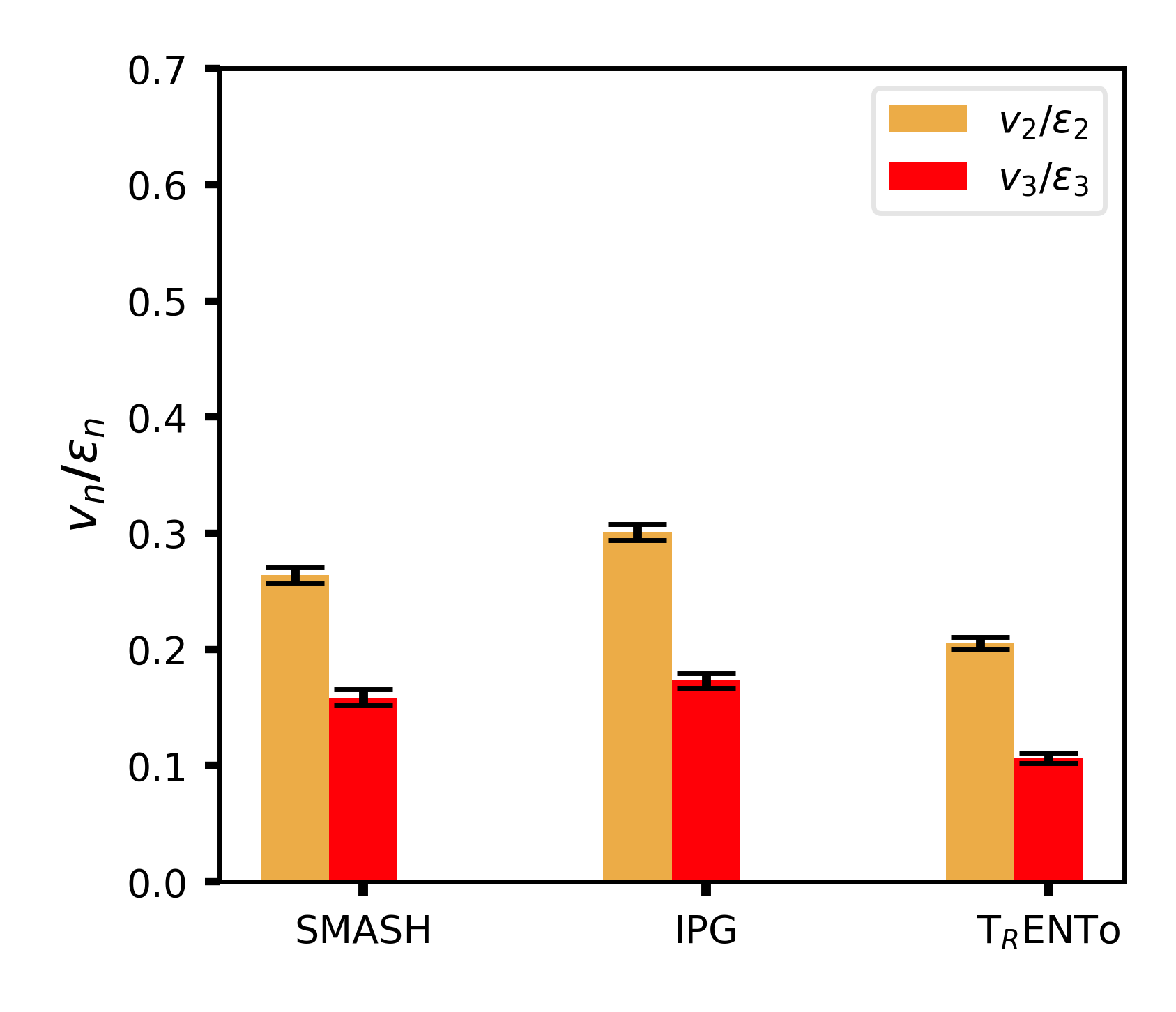}
    \caption{The response function $v_n$/$\epsilon_n$ for all the models at 20-30\% centrality.}
    \label{fig:ratio}
\end{figure}
In Figure \ref{fig:ratio}, the response for the different models is shown. One can notice a dependence on the initial condition model, although the shear viscosity parameterization and other elements of the hybrid approach are identical. Especially \trento shows a smaller response to the initial eccentricity.
The main difference between the models is the present of initial state momentum information in SMASH and IP-Glasma. From Pearson correlations alone, no other contribution to the final state flows can be identified, as can be seen in Figure
\ref{fig:corr}.
However, correlations alone can be misleading as they only capture the connection between two variables, but neglect the effect of a third independent variable. Using linear regression with final state flows as dependent variables and different sets of initial state properties as independent variables, we find consistently for both SMASH and IP-Glasma initial conditions at 0-5\% and 20-30\% centrality and both for $v_2$ and $v_3$ statistically significant improvement when performing linear regression not just with $\epsilon_n$ as independent variable, but with both $\epsilon_n$ and $\langle p_T^{IC} \rangle$. The difference in the response function for the different initial conditions can therefore be explained by the presence of further initial state factors impacting the final state flow.

%%%%%%%%%%%%%%%%%%%%%%%%%%%%%%%%%%%%%%%%%%
\section{Discussion}
This work shows the benefits of deviating from of a constant shear viscosity modifying the linear coefficient connecting $\epsilon_n$ and $v_n$. Temperature and shear viscosity dependent parameterizations of the shear viscosity can capture the change in the properties of the medium when approaching the non-equilibrium phase, and therefore mitigate the discontinuities present between the stages of a hybrid approach. Additionally, the eccentricities provided by an initial state model are not the only predictors of the final state flow. Even at comparatively high energies, the momentum space of the initial state affects the whole evolution and can significantly alter final state observables.

\vspace{6pt} 

%%%%%%%%%%%%%%%%%%%%%%%%%%%%%%%%%%%%%%%%%%
%% optional
%\supplementary{The following supporting information can be downloaded at:  \linksupplementary{s1}, Figure S1: title; Table S1: title; Video S1: title.}


%\begin{tabularx}{\textwidth}{lll}
%\toprule
%\textbf{Name} & \textbf{Type} & \textbf{Description} \\
%\midrule
%S1 & Python script (.py) & Script of python source code used in XX \\
%S2 & Text (.txt) & Script of modelling code used to make Figure X \\
%S3 & Text (.txt) & Raw data from experiment X \\
%S4 & Video (.mp4) & Video demonstrating the hardware in use \\
%\bottomrule
%\end{tabularx}
%}

%%%%%%%%%%%%%%%%%%%%%%%%%%%%%%%%%%%%%%%%%%
\authorcontributions{software, N.G. and H.E.; model selection, L.C.; computation and analysis, N.G.: writing original draft preparation, N.G.; supervision, H.E.; project administration, H.E.; funding acquisition, H.E.}

\funding{This work was supported by the Deutsche Forschungsgemeinschaft (DFG, German Research Foundation) – Project number 315477589 – TRR 211. N.G. acknowledges support by the Stiftung Polytechnische Gesellschaft Frankfurt am Main as well as the Studienstiftung des Deutschen Volkes.  Computational resources have been provided by the GreenCube at GSI. }

\conflictsofinterest{The authors declare no conflict of interest.} 

%%%%%%%%%%%%%%%%%%%%%%%%%%%%%%%%%%%%%%%%%%
%% Optional

%% Only for journal Encyclopedia
%\entrylink{The Link to this entry published on the encyclopedia platform.}

%%%%%%%%%%%%%%%%%%%%%%%%%%%%%%%%%%%%%%%%%%

%%%%%%%%%%%%%%%%%%%%%%%%%%%%%%%%%%%%%%%%%%
\begin{adjustwidth}{-\extralength}{0cm}

\reftitle{References}


%\bibliography{your_external_BibTeX_file}

%=====================================
% References, variant B: internal bibliography
%=====================================
\bibliography{hq_proceedings_ng}

% If authors have biography, please use the format below
%\section*{Short Biography of Authors}
%\bio
%{\raisebox{-0.35cm}{\includegraphics[width=3.5cm,height=5.3cm,clip,keepaspectratio]{Definitions/author1.pdf}}}
%{\textbf{Firstname Lastname} Biography of first author}
%
%\bio
%{\raisebox{-0.35cm}{\includegraphics[width=3.5cm,height=5.3cm,clip,keepaspectratio]{Definitions/author2.jpg}}}
%{\textbf{Firstname Lastname} Biography of second author}

% For the MDPI journals use author-date citation, please follow the formatting guidelines on http://www.mdpi.com/authors/references
% To cite two works by the same author: \citeauthor{ref-journal-1a} (\citeyear{ref-journal-1a}, \citeyear{ref-journal-1b}). This produces: Whittaker (1967, 1975)
% To cite two works by the same author with specific pages: \citeauthor{ref-journal-3a} (\citeyear{ref-journal-3a}, p. 328; \citeyear{ref-journal-3b}, p.475). This produces: Wong (1999, p. 328; 2000, p. 475)

%%%%%%%%%%%%%%%%%%%%%%%%%%%%%%%%%%%%%%%%%%
%% for journal Sci
%\reviewreports{\\
%Reviewer 1 comments and authors’ response\\
%Reviewer 2 comments and authors’ response\\
%Reviewer 3 comments and authors’ response
%}
%%%%%%%%%%%%%%%%%%%%%%%%%%%%%%%%%%%%%%%%%%
\PublishersNote{}
\end{adjustwidth}
\end{document}